\documentstyle[prl,aps,epsf,floats,twocolumn]{revtex}
\begin{document}
\draft
\flushbottom
\twocolumn[
\hsize\textwidth\columnwidth\hsize\csname @twocolumnfalse\endcsname

\title{Microscopic description of d-wave superconductivity by Van Hove
nesting in the Hubbard model}
\author{J. Gonz\'alez \\}
\address{
        Instituto de Estructura de la Materia.  Consejo Superior
de Investigaciones Cient{\'\i}ficas.  Serrano 123, 28006 Madrid.
Spain.}
\date{\today}
\maketitle
\begin{abstract}
We devise a computational approach to the Hubbard model that
captures the strong coupling dynamics arising when the Fermi level 
is at a Van Hove singularity in the density of states.
We rely on an approximate degeneracy among the many-body
states accounting for the main instabilities of the system
(antiferromagnetism, $d$-wave superconductivity). The Fermi line
turns out to be deformed in a manner consistent with the pinning
of the Fermi level to the Van Hove singularity. For a doping
rate $\delta \sim 0.2$, the ground state
is characterized by $d$-wave symmetry,
quasiparticles gapped only at the saddle-points of the band, and
a large peak at zero momentum in the $d$-wave pairing correlations.

\end{abstract}

]

\narrowtext 
\tightenlines


Nowadays it is widely accepted that the single-band Hubbard model
is an appropriate starting point to describe the electron
correlations in the copper-oxide materials\cite{dago}. In spite of
the efforts to show that the model has a phase of
$d$-wave superconductivity, however, there is no conclusive
proof at present of its existence. 
There have been different theoretical approaches 
to understand the symmetry of the order parameter,
among which we may quote the proposals based
on the effect of antiferromagnetic fluctuations\cite{fluc} and the
Kohn-Luttinger mechanism due to Van Hove nesting of the Fermi
surface\cite{nos,kl}. 
Numerical methods have also provided some evidence of enhanced 
pairing correlations in the 2D Hubbard model\cite{num,lad}.
Recent progress has been achieved by the use of dynamical
mean-field approximations\cite{dmf} and improved quantum Monte Carlo
methods\cite{qmc1,qmc2}.

We focus our discussion on the proposal of 
superconducting pairing by the influence of a Van Hove singularity
near the Fermi level\cite{vh}. Recently, some understanding of the
problem has been attained by the use of refined renormalization
group methods for interacting electrons in two dimensions\cite{ren,press}. 
This approach suffers, though, from the same shortcoming of any attempt
to deal with the origin of high-$T_c$ superconductivity. 
The system is likely to develop strong antiferromagnetic or
superconducting correlations but, as long as the effective
interaction grows large as more high-energy electron modes are 
integrated out, it is not
possible to discern rigorously the ground state of the model.
A related problem concerns the fact that the superconducting
correlations are enhanced like $\log^2 \varepsilon $, when the
electron modes are integrated out down to energy $\varepsilon $
near the Fermi surface. This reflects in another fashion that
the model does not lead to a conventional scale-invariant theory 
at arbitrarily low energies
---unless the Fermi energy is promoted to a dynamical quantity
susceptible itself of renormalization\cite{np,press}.

In this paper we address the characterization of the ground state
of the Hubbard model, when the Fermi level is at a Van Hove singularity
in the density of states. For this purpose we will formally make use of
the scaling properties of the model\cite{np}, which ensure that upon
integration of the high-energy modes the electron quasiparticles
remain in correspondence with the original one-particle states,
up to the energy in which the effective interaction starts to
grow large. To elucidate the behavior of the system at lower
energies we undertake a numerical diagonalization of the Hubbard
hamiltonian, in the Hilbert space truncated to include  the
states left after integration of the high-energy modes.

\begin{figure}
\epsfysize = 6cm
\centerline{\epsfbox{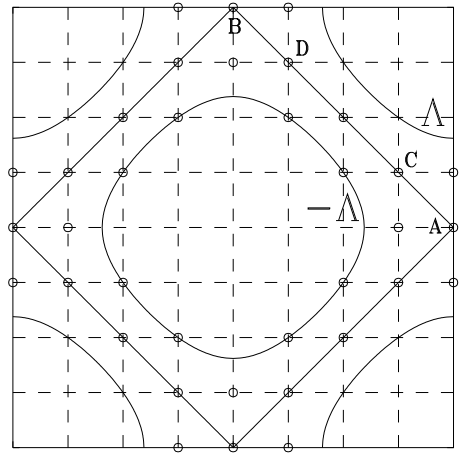}} 
\caption{Points of the $8 \times 8$ lattice in momentum space.}
\label{lattice}
\end{figure}

We implement the above strategy in the Hubbard model on a
$8 \times 8$ lattice, which has the set of one-particle states
in momentum space represented in Fig. \ref{lattice}. In order
to study the effects of the enhanced scattering at the saddle-points,
we start from a Fermi sea of occupied states below the shell
$\varepsilon = 0$, whose particle content corresponds to a doping
rate $\delta \approx 0.22$. We suppose
that at a previous renormalization group stage the high-energy
electron modes have been integrated out down to the region 
between energies $- \Lambda$ and $\Lambda $  about
the Fermi level shown in Fig. \ref{lattice}. The modes integrated
out have approximately isotropic patterns about 
the top and the bottom of the band, and we
may safely assume that their effect is encoded in a soft
renormalization of the effective coupling for the low-energy modes,
that we call $U_{\rm eff}$. Furthermore, we will deal with
a situation in which $U_{\rm eff}$ is in the weak-coupling regime, 
so that the low-energy physics can be described 
by building the many-body Hilbert
space out of the one-particle states within $- \Lambda$ and $\Lambda $.
One has to bear in mind, however, that
the interaction of the states at the Fermi level shell $\varepsilon
= 0$ is always strong and essentially nonperturbative. In fact, we
know from theoretical arguments that the main physical features have
to come from the enhanced scattering with momentum transfer
$(\pi ,\pi )$. This effect is properly captured in our construction,
given the high degeneracy of states at the $\varepsilon = 0$ shell.

We proceed then to the numerical diagonalization in the many-body
space of states of the low-energy theory, for different contents of
particles at the Fermi level shell $\varepsilon = 0$. 
We keep in all cases a Fermi sea of
24 particles filling the levels below $\varepsilon = 0$.
The instances with an even number of particles at the Fermi level
are particularly interesting, since these are situations that allow
to test the tendency to pairing in the model. Actually, there is a
correspondence between the states with broken symmetry
in a macroscopic system and the sectors labelled by different
quantum numbers in the finite cluster\cite{cluster}. 
The antiferromagnetic phase
is characterized by a ground state in the sector A
of spin singlets with total momentum
${\bf P} = (\pi ,\pi )$ and all the quantum numbers equal to 1
for the rest of discrete symmetries of the lattice. The $d$-wave
superconducting phase may be characterized by a ground state 
in the sector D of states with
total momentum ${\bf P} = (0 , 0)$ and odd parity under the exchange
of the two axes in momentum space (the rest of quantum numbers
remaining equal to 1). A ground state belonging to the sector
P of spin singlets
with total momentum ${\bf P} = (0 , 0)$ and even parity for the
rest of the symmetry operations implies a phase without
symmetry breaking at the macroscopic level.

An important observation is that, for any even number of particles, 
the lowest-energy states found in the A, D and P sectors
by considering the scattering of the one-particle states in the 
$\varepsilon = 0$ shell (and no particle-hole excitations from 
the Fermi sea) are always degenerate. Let us call ${\bf p}_A =
(\pi, 0)$, ${\bf p}_B = (0, \pi)$, ${\bf p}_C = (3\pi /4, \pi /4)$,
${\bf p}_D = (\pi /4, 3\pi /4)$. For two particles at the
Fermi level, for instance, the following states have the same
energy
\begin{eqnarray}
| A \rangle  & = &  \frac{1}{\sqrt{2}}
    \left( c^{+}_{\uparrow} ( {\bf p}_A )
            c^{+}_{ \downarrow} ( {\bf p}_B )
          -   c^{+}_{\downarrow} ( {\bf p}_A )
            c^{+}_{ \uparrow} ( {\bf p}_B )  \right)
                             | FS \rangle     \nonumber  \\
   & -  &    \frac{1}{2 \sqrt{8}} \sum_{\cal R}
    \left( c^{+}_{\uparrow} ( {\bf p}_C )
            c^{+}_{ \downarrow} ( {\bf p}_D )
          -   c^{+}_{\downarrow} ( {\bf p}_C )
            c^{+}_{ \uparrow} ( {\bf p}_D )  \right)
                             | FS \rangle     \nonumber    \\
| D \rangle  & = &    \frac{1}{\sqrt{2}}
    \left(      c^{+}_{\uparrow } ( {\bf p}_A )
               c^{+}_{\downarrow } ( {\bf p}_A  )
            -    c^{+}_{\uparrow } ( {\bf p}_B )
              c^{+}_{\downarrow } ( {\bf p}_B )   \right)
                           | FS \rangle     \nonumber    \\
| D' \rangle  & = &    \frac{1}{\sqrt{8}} \sum_{\cal R}
    \left(      c^{+}_{\uparrow } ( {\bf p}_C )
               c^{+}_{\downarrow } ( -{\bf p}_C  )
            -    c^{+}_{\uparrow } ( {\bf p}_D )
              c^{+}_{\downarrow } ( -{\bf p}_D )   \right)
                           | FS \rangle     \nonumber    \\
| P \rangle  & = &    \frac{1}{\sqrt{2}}
    \left(      c^{+}_{\uparrow } ( {\bf p}_A )
               c^{+}_{\downarrow } ( {\bf p}_A  )
            +    c^{+}_{\uparrow } ( {\bf p}_B )
              c^{+}_{\downarrow } ( {\bf p}_B )   \right)
                           | FS \rangle     \nonumber    \\
    & -  &    \frac{1}{2 \sqrt{8}} \sum_{\cal R}
    \left(      c^{+}_{\uparrow } ( {\bf p}_C )
               c^{+}_{\downarrow } ( -{\bf p}_C  )
            +    c^{+}_{\uparrow } ( {\bf p}_D )
              c^{+}_{\downarrow } ( -{\bf p}_D )   \right)
                           | FS \rangle     \nonumber 
\end{eqnarray}
where ${\cal R}$ stands for all possible reflections about the axes
and $| FS \rangle $ represents the Fermi sea below
the $\varepsilon = 0$ shell. It can be checked
that this degeneracy among the lowest-energy states in the 
three different sectors holds for any even number of particles
in the Fermi level $\varepsilon = 0$, when particle-hole excitations
from the Fermi sea are frozen out.

Thus, in order to find to which sector belongs the ground state
of the system, it suffices to make perturbations around the
zeroth-order degenerate states. Wigner theorem establishes that
no generic level crossings are found when just one parameter is varied
in a quantum system\cite{wigner}. In our problem, we may
switch adiabatically by means of an energy cutoff the number of
particle-hole processes that are allowed in the space of states,
in order to resolve the above degeneracy. The theorem guarantees
that the ground state obtained in this evolution
has to be adiabatically connected to the ground state
of the full many-body space. In practice, it is enough to perturb
the many-body states with up to two particle-hole
excitations from the Fermi sea comprising 24 particles
below $\varepsilon = 0$. At this level, we may consider our
computational method as accurate as the coupled-cluster-double
approximation, in which the relative errors estimated for the
ground state energy of the Hubbard model are as small as 
$\sim 10^{-4}$, for a $8 \times 8$ lattice and $U = t$\cite{asai}.
As long as we are going to obtain our main results from the energy
balance between different states in the model, we may expect them
to be affected by smaller relative errors at that value of 
the interaction.

We have computed the lowest-energy states in the sectors with
different quantum numbers,
within the above approximation to the space of states, for
a number $N$ of particles in the $\varepsilon = 0$ shell ranging
from zero to 4. These values are about a doping rate $\delta \sim
0.2$. Otherwise, the case $N = 4$ already sets the computational limit
in the number of particles that can be afforded with our Fermi sea.
The ground state with all the closed shells at $N = 0$ is a spin 
singlet with momentum ${\bf P} = (0 , 0)$ and even parity for the
rest of discrete symmetries. Its energy for $U_{\rm eff} = t$
turns out to be $ \approx -25.1713 t$.
The competition between the lowest energies 
in the different sectors for $N = 2$ and $N = 4$
is illustrated in Table \ref{t1}, 
for $U_{\rm eff} = t$ and $2t$. We
have found that the ground state is always in the D sector of odd
parity under the exchange of $p_x$ and $p_y$. This conclusion for
$N = 2$ can be considered a confirmation of the significance of
similar results obtained by the full numerical diagonalization
of the Hubbard model in a $4 \times 4$ cluster\cite{diag}. In the present
approach, the Fermi level at $\varepsilon = 0$ admits a larger
number of particles, and it allows us to check the trend towards
$d$-wave symmetry by filling the shell.

\begin{table}[h]
\centering
\begin{tabular}{|l|l|r|r|r|}
  & & ${\bf P} = (0,0)$ & ${\bf P} = (0,0)$ & ${\bf P} = (\pi,\pi)$ \\
  & &   $s$-wave   &   $d$-wave   &             \\  \hline
 $N = 2$ & $U = t$   & $-24.7985$   & $-24.7999$  & $-24.7985$   \\
         & $U = 2t$  & $-22.4508$   & $-22.4540$  & $-22.4507$ \\ \hline
 $N = 4$ & $U = t$   & $-24.4270$   & $-24.4273$  & $-24.4253$ \\
         & $U = 2t$  & $-21.7124$   & $-21.7159$  & $-21.7063$ \\
\end{tabular}

\caption{Lowest energies of states with different quantum numbers,
for particle content $N = 2$ and $N = 4$ in the $\varepsilon = 0$ 
shell.}
\label{t1}   
\end{table}

\begin{figure}
\epsfysize = 6cm
\centerline{\epsfbox{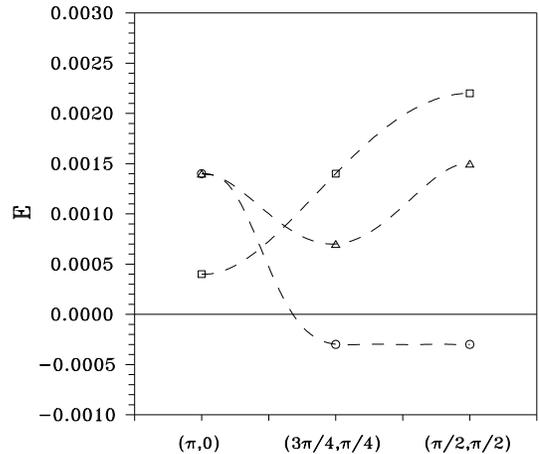}} 
\caption{Lowest energies of states with odd number of particles
$N = 1$ (squares) and $N = 3$ (triangles). The values are given
relative to the Fermi energy estimates in the text. The circles
stand for the $N = 3$ data after correction of the Fermi line
shift.}   
\label{qsp}
\end{figure}

To get insight about the nature of the ground states at $N = 2$
and $N = 4$, we have looked at the quasiparticle excitations
supported by them. These correspond in our cluster to states with
an odd number of particles ($N = 1$ and $N = 3$).
States with $N = 1$ can be considered
quasiparticle excitations over the fully symmetric state with
$N = 0$, and their dispersion may be obtained by looking at
different momenta above the Fermi sea. We have plotted in Fig.
\ref{qsp} the lowest energies found for $U_{\rm eff} = t$
among the many-body states with different momenta on the shell 
$\varepsilon = 0$ (including again up to two particle-hole
excitations from the Fermi sea).

The Fermi energy for $N = 1$ can be estimated by taking the mean
value between the ground state energies at $N = 0$ and $N = 2$,
which turns out to be $\approx - 24.9856 t$ for $U_{\rm eff} =
t$. The lowest energy found at ${\bf P} = (\pi , 0)$ is slightly
above this value, by an amount of order $10^{-4} t$. The
quasiparticle excitations at ${\bf P} = (\frac{3\pi}{4} ,
 \frac{\pi}{4})$ and ${\bf P} = (\frac{\pi}{2} , \frac{\pi}{2})$
are found at higher energies. Results pointing in the same direction have 
been obtained in the diagonalization of small clusters\cite{qp1}.
We find a clear indication that
the Fermi line is deformed by the interaction, in such a way that
it is shifted inwards at the momenta $(\frac{3\pi}{4} ,
 \frac{\pi}{4})$ and $(\frac{\pi}{2} , \frac{\pi}{2})$\cite{qp2}. This
deformation is in agreement with the effect of pinning of the
Fermi energy near a Van Hove singularity that has been discussed
on theoretical grounds\cite{mark,nos}. 
According to this effect, the particles
take advantage of the strong screening of the interaction at the
singularity to place themselves preferently near the saddle points
of the dispersion relation, as observed in our computation.
A similar phenomenon has been advocated recently in Ref.
\onlinecite{ogata}.

It is also of great interest the study of the lowest-energy states
for $N = 3$ particles on the $\varepsilon = 0$ shell. These may be
considered as the quasiparticle excitations of the ground states
for $N = 2$ and $N = 4$, and they are essential to characterize the
pairing of particles at even filling levels. The
energies for the momenta on the $\varepsilon = 0$ shell, obtained 
in the same approximation as before, are plotted in Fig. \ref{qsp}
for $U_{\rm eff} = t$.

The position of the
Fermi energy at $N = 3$  can be estimated from the ground state
energies at $N = 2$ and $N = 4$, and it becomes $\approx -24.6136 t $
for $U_{\rm eff} = t$. The deviation from this value of the
lowest-energy states for the odd-filled shell may signal the momenta 
at which a gap opens up for the quasiparticle excitations.
Given the above mentioned shift of the Fermi line, though, one has
to bear in mind that the Fermi level is not reached precisely at
the points ${\bf P} = (\frac{3\pi}{4} , \frac{\pi}{4})$ and
${\bf P} = (\frac{\pi}{2} , \frac{\pi}{2})$. This effect can be
corrected by subtracting
from the respective energies the shift from the Fermi level as
measured for the $N = 1$ quasiparticles. The final result can be seen
in Fig. \ref{qsp}. The energy of the quasiparticles near
${\bf P} = (\frac{3\pi}{4} , \frac{\pi}{4})$ and
${\bf P} = (\frac{\pi}{2} , \frac{\pi}{2})$ turns out to be
the same, within
the precision of our computation, and slightly below the estimate
from the mean value of the ground state energies for $N = 2$ and
$N = 4$. On the contrary, the quasiparticle at ${\bf P} = (\pi , 0)$
has an energy well above the Fermi energy estimate. It is clear that
a gap opens up for quasiparticle excitations, but only at
the position of the saddle points. The value $\Delta $ of the gap
for $U_{\rm eff} = t$, $\Delta \approx 0.0017 t$, is sensibly greater
than the energy difference between the lowest-energy states with $s$-wave
and $d$-wave symmetry for $N = 4$.

The conclusive evidence that the $d$-wave ground state at $N = 4$
is built out of pairing of particles is given by the measure of
pairing correlations with $d$-wave symmetry. We have computed the
correlations of the operator
\begin{equation}
O_{SCD} ({\bf q}) = \sum_{\bf k} \left(\frac{ g({\bf k}+{\bf q}) +
   g({\bf k}) }{2}  \right)
c^{+}_{\uparrow} ({\bf k}+{\bf q}) c^{+}_{\downarrow} (-{\bf k})
\end{equation}
with $g({\bf p}) = \cos (p_x) - \cos (p_y)$, in the ground state 
for $N = 4$ at $U_{\rm eff} = t$. The results are shown in Table 
\ref{t2}. The momenta that can be measured are enough to 
discern a high peak at ${\bf q} = 0$ of the pairing correlations,
together with a sharp decrease away from the origin.
To establish some comparison, we may
quote that the peak of the $s$-wave pairing correlations in the 
lowest-energy state of the P sector is $\approx 6.9$. Taking
into account that the $d$-wave correlations decay by one order of
magnitude across two lattice units in momentum space, the translation
of the results to real space is that the binding of particles is
correlated on the whole dimension of the cluster.
 

\begin{table}[h]
\centering
\begin{tabular}{|c|r|}
  ${\bf q}$   & $\langle O_{SCD} ({\bf q}) O_{SCD}^{+} ({\bf -q})  
              \rangle $  \\
 \hline
  $(0,0)$     &    $ 25.969 $     \\
  $(\frac{\pi}{4},0)$                   &   $ 4.907 $    \\
  $(\frac{\pi}{4},\frac{\pi}{4})$     &    $ 8.526 $  \\
  $(\frac{\pi}{2},0)$                   &  $ 3.262 $     \\
  $(\frac{\pi}{2},\frac{\pi}{4})$     &   $ 1.671 $   \\
  $(\frac{\pi}{2},\frac{\pi}{2})$     &  $ 1.059 $    \\
\end{tabular}

\caption{Values of the pairing correlator for different
momenta near the origin.}
\label{t2}
\end{table}

Although our results refer to a $8 \times 8$ cluster, 
they are robust enough to survive the extrapolation to larger lattices.
The reason for that is the high degree of symmetry of the model before
switching on the interaction with the Fermi sea. We have seen that,
by focusing on the scattering processes at the Fermi level shell
$\varepsilon = 0$, the lowest-energy state with momentum 
${\bf P} = (\pi , \pi)$ and those with $d$-wave and $s$-wave symmetry 
turn out to be degenerate. This is a universal feature
of the interaction of states on the nested Fermi line, irrespective
of lattice size. Our approach to the many-body problem starts by
solving first the strong interaction in the 
highly degenerate Fermi level. While the interaction with the rest of
excitations from the Fermi sea is relevant, as it resolves the
mentioned degeneracy, its effect may be assessed within our 
weak-coupling renormalization group scheme. 

The existence of the $d$-wave ground state with remarkable pairing
correlations provides a microscopic basis for a long-standing
conjecture. There has been much hope that a mechanism based on
antiferromagnetic fluctuations may be the source of superconductivity
in the copper-oxide materials. 
A different approach ---though similar in essence--- has put forward the
idea that the Hubbard model should bear a form of Kohn-Luttinger
superconductivity near half-filling.
By stressing the relevance of the
modes near the Fermi level shell at $\varepsilon = 0$, our explicit
construction of the $d$-wave ground state 
points at the processes with momentum transfer ${\bf Q} = 
(\pi , \pi)$ as the source of an attractive channel in the Hubbard 
model with bare repulsive interaction. 
Nesting of the saddle points at the Fermi level is at the origin
of the relevance of such processes, which single out $d$-wave
superconductivity as the dominant instability of the system.

I thank F. Guinea for useful comments on the manuscript.
This work has been partially supported by the spanish Ministerio de
Educaci\'on y Cultura Grant PB96-0875.

\end{document}